\documentstyle[12pt]{article}

\begin{document}
\title{The Effect of Sources on the Inner Horizon of Black Holes }
\author{ Ozay Gurtug and Mustafa Halilsoy \\
Department of Physics, Eastern Mediterranean University \\
G.Magusa, North Cyprus, Mersin 10 - Turkey \\
email: ozay.gurtug@emu.edu.tr}
\maketitle
\begin{abstract}
{\small Single pulse of null dust and colliding null dusts both
transform a regular horizon into a space-like singularity in the
space of colliding waves. The local isometry between such
space-times and black holes extrapolates these results to the
realm of black holes. However, inclusion of particular scalar
fields instead of null dusts creates null singularities rather
than space-like ones on the inner horizons of black holes.}
\end{abstract}
\newpage
\section{Introduction}
In this paper we show, with {\bf exact} solutions, the Cauchy
horizon (CH) has indeterminate character with respect to different
perturbing potentials and in some cases the CH can be space-like
singular and in others it can be null singular. \\
The first signs of this sort of indeterminacy of character for CH
were seen in the outcomes of Chandrasekhar and Xanthopoulos (CX)
[1] and Yurtsever's [2] analysis of the stability of the horizon
(or quasiregular singularity) of the Bell-Szekeres (BS) [3]
space-time. These authors used
perturbation methods in their work.\\
In Ref. [4] it has been shown there is a similar inner horizon
instability for black holes (BHs) and the horizons change to
space-like singularities. On the other hand, Ori [5] found the
horizon of a Kerr BH changes to a null
singularity. All of these works used perturbation methods. \\
This lack of consensus for the instability of colliding plane
waves (CPW) and BH horizons attracted much interest and effort.
Burko [6,7] confirmed Ori's findings of a regular horizon changing
to a null singularity when he applied a scalar field to a
Reissner-Nordstrom (RN) BH. His work was done using numerical
methods. \\
As an alternative to Burko's approach we have applied the local
isometry between CPW and the region between the two (Event and
Cauchy) horizons of BHs. This enables us to couple scalar and
other fields to CPWs, where there are analytically tractable
solutions, and then to transform our results to BH cases. This
approach was first introduced by Yurtsever [8]. Yurtsever
concluded that the instabilities of the CHs in Kerr and RN BHs
turns the CH to a space-like curvature singularity. Yurtsever's
comments were indications of certain possibilities but they
remained as
unsupported comments. \\
We exploit the isometry analogy and we consider two sorts of
sources in the CX CPW [9] space-time. The first is null
propagating dust ( with their mutual collision ). We concentrate
our attention on the CX space-time because it has a nonsingular
horizon and it is locally isometric to the Kerr-Newman (KN) BH,
enabling us to transform through isometry to the BH space-time. We
find that propagation (or collision) of null dusts in the CX
space-time convert the inner horizon to a space-like singularity.
This conclusion is supported by CX [10] for the Einstein-Vacuum
problem which is locally isometric to the Kerr BH. CX's solution
supports our conclusion applied to the KN BH, because the presence
of charge is trivial. We point out that the introduction of
coupling in our case is entirely different from
the one in CX. \\
The second is a scalar field in between the horizons. We show that
such a field effects the inner horizon differently from the above
case and the singularity it creates turns out to be null. In other
words our two sorts of coupling show the inner horizon of a BH
does not have a unique character in its singularity structure and
this character depends on the perturbing potential. The
instabilities of these CHs occuring in CPW space-times and those
of
corresponding BH space-times have dual character. \\
 We also make some comments about the  Helliwell - Konkowski (HK) [11]
  conjecture. HK conjecture was thought to enable us to predict the instability of a
 horizon and the sort of singularity it changes to. However, our
 conclusion that inner horizons have dual character shows that HK
  conjecture cannot uniquely determine the sort of the outcoming singularity,
 hence it should be used with caution. \\
This paper is organized as follows. Section 2 we review the
connection between the CX and the KN space-times. In section 3 we
consider null dust as a test field in the CX space-time. Section 4
follows with an exact back-reaction solution to the foregoing
section. Section 5 exposes the role of scalar fields leading to
the null singularities in the KN BH space-time. We conclude the
paper with a discussion in section six.
\section{Chandrasekhar-Xanthopoulos and Kerr - Newman  Metrics}
  CX have found a colliding wave solution in the Einstein - Maxwell (EM) theory which
  is locally isometric to the KN BH solution [9]. CX metric is
  given by
\begin{equation}
ds^2=X\left(\frac{d\tau^2}{\Delta}-\frac{d\sigma^2}{\delta}\right)
-\Delta\delta \frac{X}{Y} dy^2 - \frac{Y}{X}\left(dx-q_{2e}dy\right)^2
\end{equation}
where the coordinates $(\tau,\sigma)$ are given in terms of the null coordinates
$(u,v)$ by
\begin{eqnarray}
\tau&=&u\sqrt{1-v^2}+v\sqrt{1-u^2} \nonumber \\
\sigma&=&u\sqrt{1-v^2}-v\sqrt{1-u^2} \nonumber
\end{eqnarray}
and $ \Delta = 1- \tau^2 \, , \delta = 1 - \sigma ^2 $. \\
The metric functions are
\begin{eqnarray}
X&=&\frac{1}{\alpha^2}\left[(1-\alpha p \tau)^2+\alpha^2q^2\sigma^2\right] \nonumber \\
Y&=&1-p^2\tau^2-q^2\sigma^2 \nonumber \\
q_{2e}&=&-\frac{q\delta}{p\alpha^2} \frac{1+\alpha^2-2\alpha p \tau}{1-p^2\tau^2-q^2\sigma^2}
\end{eqnarray}
in which the constants $\alpha,p$ and $q$ must satisfy
\begin{eqnarray}
0<\alpha\leq1 \nonumber \\
p^2+q^2=1
\end{eqnarray}
The metric (1) transforms into the Boyer-Lindquist form of the KN, if the following
transformation is used.
\begin{equation}
t=m \alpha x, \hspace{.5cm} y=p\phi, \hspace{.5cm} \tau=\frac{m-r}{\sqrt{m^2-a^2-Q^2}}, \hspace{.5cm}
\sigma=\cos\theta
\end{equation}
with
\begin{equation}
p=\frac{\sqrt{m^2-a^2-Q^2}}{m\alpha}, \hspace{1cm} q=-\frac{a}{m\alpha}, \hspace{1cm}
(m^2>a^2+Q^2)
\end{equation}
so that $Q^{2}=(1-\alpha^{2})m^{2}$ holds. Note that $ \alpha = 1 $ removes
the charge
and reduces the problem from KN to Kerr and in particular the limit $ a=0 $
yields the  Reissner - Nordstrom case.
With these substitutions the line element (1) may be written in the form \\
\begin{eqnarray}
\alpha ^2 m^2 ds^2 &=& \left( 1-\rho ^{-2} \omega \right) dt^2 - \sin ^2 \theta \left[ \Delta
+ \omega \left(1+ a^2 \rho ^{-2} \sin^2 \theta \right) \right] d\phi^2 \nonumber \\
& & \nonumber \\
& &- 2 a \omega \rho^{-2} \sin^2 \theta dt d\phi - \rho^2 \left( \Delta^{-1} dr^2 + d \theta^2 \right)
\end{eqnarray}
with the standard notations
\begin{eqnarray}
\Delta&=&r^2-2mr+a^2+Q^2\equiv(r-r_{-})(r-r_{+}) \nonumber \\
\rho^2&=&r^2 + a^2\cos^2\theta \nonumber \\
\omega&=&2mr-Q^2 \nonumber
\end{eqnarray}
in which $a$ and $Q$ stand respectively for the constants of rotation and electric
charge. Note that $ \Delta $ here is different from the $ \Delta $ of the CX metric.
 The roots of $\Delta,r_{+}$ and $r_{-}$ are known as the Event and
Cauchy (inner) horizon, respectively. Therefore the colliding wave solution due to
CX is locally isometric to the KN metric in between the two horizons.

\section{Test Null Dusts in the CX Spacetime}
We consider now two null test dusts moving in opposite directions
in the interaction region of the CX metric or equivalently null
fields moving in the isometric region of the KN space-time. Such
null dusts (and the following exact solution) suffice to expose
the non-linear effect of the background as well as the disturbance
of the background (i.e the back-reaction). This is provided by
appealing to the null geodesics of the KN and transforming back
via (4) and (5) to the CX metric. For simplicity we choose
$\sigma=0$ in CX (or $\theta=\pi/2 $ in KN) to obtain the first
integrals of the null geodesics as
\begin{equation}
\dot{t}=\frac{E(r^2+a^2)}{\Delta}, \hspace{1cm} \dot{\phi}=-\frac{aE}{\Delta},
 \hspace{1cm} \dot{r}=E
\end{equation}
in the KN geometry, and the corresponding first integrals of the CX geometry are
\begin{eqnarray}
\dot{\tau}=\frac{E}{m\alpha p}, \hspace{1cm} \dot{y}=\frac{qE}{m\alpha p(1-\tau^2)} \hspace{3cm} \nonumber \\
\dot{x}=-\frac{E}{m^3\alpha^3 p^2 (1-\tau^2)} \left\{ m^2\left[(1-\alpha p \tau )^2+
(1-\alpha^2 p^2) \right] -Q^2 \right\}
\end{eqnarray}
In both cases $E$ is the energy constant and dot represents the appropriate parameter
for the null geodesics. We insert two null dust congruences with finite densities
$\rho_{l}$ and $\rho_{n}$ propagating along the null vectors $l_{\mu}$ and $n_{\mu}$.
In other words the total test energy-momentum tensor is
\begin{equation}
T_{\mu\nu}=\rho_{l}l_{\mu}l_{\nu}+ \rho_{n}n_{\mu}n_{\nu}
\end{equation}
where
\begin{eqnarray}
l^\mu&=&\left(1,-\frac{\left[(1-\alpha p\tau)^2+\alpha^2q^2\right]}{\alpha^2
p(1-\tau^2)},\frac{q}{1-\tau^2},0\right) \nonumber \\
& & \nonumber \\
n^\mu&=&\left(1,\frac{\left[(1-\alpha
p\tau)^2+\alpha^2q^2\right]}{\alpha^2
p(1-\tau^2)},-\frac{q}{1-\tau^2},0\right)
\end{eqnarray}
in which we have scaled
\begin{equation}
\frac{E}{m\alpha p}=1  \nonumber
\end{equation}
The non-trivial scalar $T_{\mu \nu}T^{\mu \nu}$ of the criss-crossing null test dust
is given by
\begin{equation}
 T_{\mu\nu}T^{\mu\nu}=8\rho_{l}\rho_{n}\frac{(1-\alpha p \tau)^4}{\alpha^4(1-\tau^2)^2}
\end{equation}
 which diverges for $ \tau \rightarrow 1 $. This corresponds to a divergence for $r\rightarrow r_{-}$
 in the KN black-hole. As a prediction of the HK conjecture any exact back-reaction
 solution that is imitated by the foregoing test dust must destroy the horizon.
  In the next section we present a new exact back-reaction
 solution which represents a solution of colliding Einstein-Maxwell-Null dust that exhibits
 a space-like singularity  for $\tau\rightarrow 1$. The new solution incorporates
a regular conformal factor (such that it does not diverge as $
\tau \rightarrow 1 $ ) and therefore leaves all Weyl scalars
invariant and regular.
 \section{A New Exact Back-reaction Solution}
 Our aim now is to present an exact solution which involves collision of
 Einstein-Maxwell fields coupled with null shells. The shells are added as a conformal
 factor and our method can be summarized as follows. \\
 Let $ds_{0}^2$ represent the CX metric (1) which is isometric to the KN. Then, the
 new metric [12]
 \begin{equation}
 ds^2=\frac{1}{\phi^2}ds_{0}^2
 \end{equation}
 where $\phi=1+\alpha_{0}u\theta(u)+\beta_{0}v\theta(v)$, with $(\alpha_{0},\beta_{0})$
 positive constants and $\theta$ standing for the step function, represents colliding
 Einstein-Maxwell fields coupled with null shells. This metric has some advantages
 over the back-reaction solution of CX. First, while Ricci components and scalar curvature
 (if any) are affected by inclusion of the conformal factor the Weyl scalars remain
 invariant i.e they are finite on the horizon. It turns out as shown in
 Appendix A explicitly that the scalar curvature and some Ricci components diverge on the horizon.
 Thus it is misleading to judge the behaviour of a horizon by looking only at the Weyl
 scalars. Our approach gives the clue: it is more reliable to investigate the behaviours
 of the scalar curvature and the Ricci components. In this sense the solution adopted
 in (13) as the exact version of the test null dust is stronger (and simpler)
 than the implication of the source added CX solution [9]. The
 acceptability of the inclusion of this conformal factor is
 shown by checking the weak and dominant energy conditions of
 the new solution. The details are given in Appendix B. \\
 As a second advantage we point out that the metric
 \begin{equation}
 ds^2=(1+\alpha_{0}u +\beta_{0}v)^{-2}\left(2dudv-dx^2-dy^2\right)
 \end{equation}
 represents the de Sitter space with scalar curvature and cosmological constant
 as the only non-zero physical quantities [13]. The transformation
\begin{eqnarray}
1+\alpha_{0}u +\beta_{0}v&=&e^{\lambda t} \nonumber \\
  \alpha_{0}u -\beta_{0}v&=&\lambda z
\end{eqnarray}
takes this metric into
\begin{equation}
ds^2=dt^2-e^{-2 \lambda t} \left( dx^2 + dy^2 + dz^2 \right)
\end{equation}
which is identified as the de Sitter metric. Similarly by the choice of the conformal
factor and transformation
\begin{eqnarray}
1+\alpha_{0}u -\beta_{0}v&=& \lambda z \nonumber \\
  \alpha_{0}u +\beta_{0}v&=&\lambda t
\end{eqnarray}
our metric becomes
\begin{equation}
ds^2= \frac{1}{\lambda^2 z^2} \left( dt^2 - dx^2 - dy^2 - dz^2 \right)
\end{equation}
which is the anti - de Sitter metric. In both cases the constant $
\lambda $ is defined by $ \lambda = \sqrt{2 \alpha_{0} \beta_{0} }
$ in which $ \alpha_{0}>0 , \beta_{0}>0 $. Now instead of the flat
metric by substituting the CX metric (1) it can be interpreted as
colliding Einstein - Maxwell fields in a de Sitter background.
This is the alternative interpretation that our metric admits when
we remove the step functions in the conformal factor. By this
interpretation (and of course through the isometry) it says that a
KN BH endowed with a de Sitter background in between the horizons
gives rise to a divergent scalar curvature besides some of the
Ricci's. In turn according  the HK conjecture such a horizon
converts into a singularity. This singularity is necessarily
space-like since a normal vector to the horizon turns out to be
time-like. As a final advantage we recall that in the CX metric
the null fields are added to the vacuum problem whereas in our
case we make the addition to the electrovacuum problem. Such an
extension was a missing link in the study of CX.
\section{Null Singularities in the Presence of Scalar Fields}
Recently, we have shown the existence of null singularities in the
CPW space-time
for a class of linearly polarized metrics [14]. \\
 Our main objective in this section is to investigate the
singularity structure of the CHs that exist in  charged spinning
BHs, namely KN BH in the presence of scalar fields. This is
achieved through the existing isometry between the CPW
space-time  and the BHs space-time. \\
The adopted space-time line element to describe the collision of
plane waves with non-parallel polarization is given by
\begin{equation}
ds^2=2e^{-M}dudv-e^{-U}\left[ \left(e^V dx^2 +e^{-V}dy^2 \right)
\cosh W - 2 \sinh Wdxdy  \right ]
\end{equation}
The complete set of partial differential equations for the metric
functions is given elsewhere [15]. The $u,v$ dependent massless
scalar field equation
\begin{equation}
\partial _{\mu} \left ( g^{\mu \nu} \sqrt{g} \phi _{\nu} \right)=0
\end{equation}
reads as
\begin{equation}
2\phi _{uv}=U_{u} \phi _{v} + U_{v} \phi_{u}
\end{equation}
where $ \phi $ is the scalar field. Given any EM solution we can
generate an Einstein-Maxwell-Scalar (EMS) solution in accordance
with the shift.
\begin{equation}
M \rightarrow \tilde{M} + \Gamma
\end{equation}
Where $ \tilde{M} $ is any EM solution and
\begin{eqnarray}
\Gamma _{u} U_{u}=2 \phi^2_{u} \nonumber \\
\Gamma _{v} U_{v}=2 \phi^2_{v}
\end{eqnarray}
The integrability conditions for the latter equations implies that
\begin{equation}
\left( \phi_{u}U_{v}-\phi_{v}U_{u} \right)\left(2\phi _{uv}-U_{u}
\phi _{v} - U_{v} \phi_{u}\right)=0
\end{equation}
Now any solution of the scalar equation (21) helps us to construct
the extra metric function $ \Gamma $ by the line integral
\begin{equation}
\Gamma= 2 \int \frac{ \phi^2_{u}}{U_{u}} du + 2 \int \frac{
\phi^2_{v}}{U_{v}} dv
\end{equation}
As an EM solution, we use the metric (1) obtained by CX which was
shown to be locally isometric to the KN BH that uses $ \tau ,
\sigma $ coordinates instead of $ u,v $. \\
In terms of $ \tau , \sigma $ the scalar field equation (21) and
the condition (23) are equivalent to
\begin{equation}
\left[ ( 1- \tau ^2) \phi_{ \tau} \right]_{ \tau} - \left[ (1-
\sigma ^2 ) \phi_{ \sigma} \right] _{ \sigma}=0
\end{equation}
and
\begin{eqnarray}
(\tau ^2- \sigma ^2) \Gamma_{\tau }&=&2 \Delta \delta \left( \tau
\phi^2 _{\tau} + \frac{\tau \delta}{\Delta} \phi^2_{\sigma} -2
\sigma \phi_{\tau} \phi_{\sigma} \right) \nonumber \\
(\sigma ^2- \tau ^2) \Gamma_{\sigma }&=&2 \Delta \delta \left(
\sigma \phi^2 _{\sigma} + \frac{\sigma \Delta}{\delta}
\phi^2_{\tau} -2 \tau \phi_{\tau} \phi_{\sigma} \right)
\end{eqnarray}
respectively. For the present problem we choose the scalar field $
\phi $ as
\begin{equation}
\phi= \frac{k_{1}}{2} \ln{ \frac{1+ \tau}{1- \tau}} +
\frac{k_{2}}{2} \ln{ \frac{1+ \sigma}{1- \sigma }}
\end{equation}
in which $ k_{1} $ and $ k_{2} $ are constant parameters. Using
the equations (27), the metric function $ \Gamma $, due to the
scalar field is found as
\begin{equation}
e^{\Gamma}= \Delta ^{-k^2_{1}} \delta ^{-k^2_{2}}\left( \tau +
\sigma \right)^{(k_{1} + k_{2} )^2 }\left( \tau - \sigma
\right)^{(k_{1}-k_{2})^2}
\end{equation}
The new metric that describes the collision of EMS fields is
expressed by
\begin{equation}
ds^2=Xe^{-\Gamma}\left(\frac{d\tau^2}{\Delta}-\frac{d\sigma^2}{\delta}\right)
-\Delta\delta \frac{X}{Y} dy^2 -
\frac{Y}{X}\left(dx-q_{2e}dy\right)^2
\end{equation}
Where $ X,Y, \Delta , \delta $ and $ q_{2e} $ are given in
equation (2). It is well known that the CX solution is a regular
solution. However, coupling a scalar field $ \phi $, transforms
the CH into a scalar curvature singularity (SCS). This can be seen
from the scalar,
\begin{equation}
4 \pi T^{\mu}_{\mu}= \frac{e^{\Gamma}}{\Delta \delta X}
\left(\Delta k^2_{2}- \delta k^2_{1} \right)
\end{equation}
that as $ \tau \rightarrow 1 $, it becomes divergent; and hence it
is a SCS. The interesting property of this new solution is that,
the type of the singularity is null rather than space-like. This
can be justified as follows \\
The singular (or horizon) surface is described by
\begin{equation}
S(\tau)=1-\tau \nonumber
\end{equation}
(in which we have assumed an equatorial plane namely $ \sigma =0
$, in the transformed BHs space-time ).\\
We compute the normal vector to this surface
\begin{equation}
( \nabla S )^2=g^{\tau \tau } S^2_{\tau
}=\frac{(1-\tau^2)e^{\Gamma}}{X}=\frac { \alpha ^2 \tau
^{2(k^2_{1}+k^2_{2})}(1- \tau ^2 )^{1-k^2_{1}}}{(1-p \alpha \tau
)^2}
\end{equation}
Since we are interested in the limit as $ \tau \rightarrow 1 $,
for $ k^2_{1} <1 , ( \nabla S )^2=0 $ which indicates a null
property. The blow-up of the curvature scalar provides a generic
null singularity on this surface. The existing isometry can
provide us a null singularity in the corresponding  BH problem.\\
Using the transformations given in equations (4) and (5) the
metric (30) transforms into
\begin{eqnarray}
\alpha ^2 m^2 ds^2 &=& \left( 1-\rho ^{-2} \omega \right) dt^2 -
\sin ^2 \theta \left[ \Delta
+ \omega \left(1+ a^2 \rho ^{-2} \sin^2 \theta \right) \right] d\phi^2 \nonumber \\
& & \nonumber \\
& &- 2 a \omega \rho^{-2} \sin^2 \theta dt d\phi - \rho^2
e^{-\Gamma }\left( \Delta^{-1} dr^2 + d \theta^2 \right)
\end{eqnarray}
where $ \rho , \omega $ and $ \Delta $ are given in (6). This
metric represents the KN BH coupled with the particular scalar
field which reads in the equatorial plane
\begin{equation}
\phi=\frac{k_{1}}{2}\ln{ \left(\frac{r_{+}-r}{r-r_{-}} \right)}
\end{equation}
In the same plane we have
\begin{equation}
e^{\Gamma}=\left[ \left(r_{+}-r \right) \left(r-r_{-} \right)
\right]^{-k^2_{1}} \left( \frac{m-r}{\sqrt{m^2-Q^2}}
\right)^{2(k^2_{1}+k^2_{2})}
\end{equation}
Similar analysis on the singular surface,
\begin{equation}
S(r)=r-r_{-} \nonumber
\end{equation}
results in $ \left( \nabla S \right)^2=0 $ as $ r \rightarrow
r_{-} $. This result retains the null singularity formation on the
corresponding BH problem.\\
Note that, the scalar field in equation (28) is singular as $ \tau
\rightarrow 1 $. Let us now choose the following scalar field
which is regular as $ \tau \rightarrow 1 $.
\begin{equation}
\phi = \alpha _{0} \tau \sigma
\end{equation}
where $ \alpha _{0} $ is any constant parameter. The metric
function $ \Gamma $ is obtained as
\begin{equation}
e^{ \Gamma }=e^{\alpha _{0} [ \tau^2 + \sigma ^2 (1- \tau ^2)]}
\end{equation}
The curvature scalar due to this scalar field is
\begin{equation}
4 \pi T^{\mu}_{\mu}= \frac{\alpha ^2_{0} e^{\Gamma}}{X} \left(
\tau ^2 \delta - \sigma ^2 \Delta \right)
\end{equation}
It is clear to see that, the space-time remains regular as $ \tau
\rightarrow 1 $.
\section{Discussion}
It is known that the relations between the mathematical theory of
BHs and that of CPW space-time requires that, the interior of
every standard BH solution is locally isometric to the interaction
region of CPW space-time. \\
In this paper, we have used this fact to investigate the
singularity structure of the charged spinning BH namely the KN. In
our analysis, we have coupled matter (null shells) and scalar
fields to the geometry of CPW space-time and using the isometry,
we transformed the  resulting metric to the BH geometry and
investigated the effect of these fields on the CH. \\
Since our analysis is based on a completely analytic exact
solutions, we believe that, it reflects the real character of the
Killing
CHs of CPW and BH space-times. \\
First, a pair of test null dust is inserted in the CX colliding
wave space-time. Since the energy - momentum scalar diverges then
we conclude that the CH is unstable and an exact back reaction
solution must yield a scalar curvature singularity. The exact
solution which we present involves a regular conformal factor as a
source so that all Weyl scalars remain finite while some Ricci's
(and scalar curvature) diverge on the horizon. Inclusion of the
conformal factor is equivalent energetically to the de Sitter
background and through the isometry it makes possible for us to
embed a BH in a de Sitter (or anti - de Sitter) background.
Although we have inserted a matter field as a regular conformal
factor, the slightest effect of the source (through the scalar
curvature or cosmological constant ) of such a background suffices
to destroy the inner horizon and converts it into a space-like
curvature singularity.\\
Second, we have inserted  scalar fields to the geometry of CX CPW
solution. Inclusion of particular scalar fields was shown to
create null singularities rather than space-like ones both in the
space of colliding waves and that of corresponding BH space-times.
With these particular exact solutions, we may conclude that the
Killing CH of CPW space-times and the inner horizon of BHs have
dual character when they are subjected to the inclusion of matter
and
scalar fields. \\
It remains an open question whether this null singularity is an
intermediate stage between the horizon and space-like curvature
singularity.
\section*{Appendix: A \\ The Ricci and Curvature Scalars}
The non-zero Ricci and curvature scalars of the collision of Einstein - Maxwell
fields coupled with null shells ( in the same null tetrad employed by CX )
 are found as follows.
\begin{eqnarray}
\Phi_{22}&=&(\Phi_{22})_{CX} + \alpha _{0} \frac{e^M}{\phi} \left[
\delta (u) + \theta
 (u) M_{v} \right] \\
\Phi_{00}&=&(\Phi_{00})_{CX} + \beta _{0} \frac{e^M}{\phi} \left[
\delta (v) + \theta
 (v) M_{u} \right] \\
\Phi_{02}&=&(\Phi_{02})_{CX} - \frac{e^M}{2 \phi} \left \{ \alpha
_{0} \left[ V_{v} \cosh W - i W_{v} \right]
\theta (u) \right. \nonumber \\
& & \nonumber \\
& & \left. + \beta _{0} \left[ V_{u} \cosh W - i W_{u} \right] \theta (v) \right \} \\
\Phi_{11}&=& \frac{e^M}{2 \phi (1-u^2-v^2)} \left[ \beta _{0} u  + \alpha _{0} v  \right] \theta (u) \theta (v) \\
& & \nonumber \\
\Lambda &=& \frac{e^M}{12 \phi} \left \{ \frac{ \left[ \beta _{0}u
 + \alpha _{0} v  \right] } {(1-u^2-v^2)} + 12 \alpha _{0} \beta
 _{0}
\phi^{-1}  \right \}\theta (u) \theta (v)
\end{eqnarray}
In these expressions $ ( \Phi_{22})_{CX}, ( \Phi_{00})_{CX} $ and $ ( \Phi_{02})_{CX} $
refer to the CX quantities which are all finite on the horizon. Similarly all
Weyl scalars are same as in the CX metric, namely regular on the horizon. The
other functions $ M, V $ and $ W $ are given by the following expressions.
\begin{eqnarray}
e^V \cosh W &=& \frac{Y}{X \sqrt{\Delta \delta}} \nonumber \\
e^{-V} \cosh W &=& \frac{X \sqrt{\Delta \delta}}{Y} + \frac{Y q^2_{2e}}{X \sqrt{\Delta \delta}} \nonumber \\
e^{-M} &=& \frac{2X}{\sqrt{1-u^2} \sqrt{1-v^2}} \nonumber \\
\phi &=& 1 + \alpha _{0} u \theta(u) + \beta _{0} v \theta(v)
\end{eqnarray}
Since $ (1-u^2-v^2)=\sqrt{1-\tau ^2} \sqrt{1-\sigma ^2} $ for $
\sigma = 0 $ and $ \tau \rightarrow 1 $ (on the horizon) the
divergence of the scalar curvature $ \Lambda $ and $ \Phi_{11} $
is clearly manifest. A detailed calculation reveals that $
\Phi_{02} $ is also divergent while $ \Phi_{00} $ and $ \Phi_{22}
$ remain finite on the inner horizon. Let us note that suppressing
one of the incoming shells but retaining the other, still the
above components diverge. This amounts to colliding Einstein -
Maxwell wave from one side with an Einstein - Maxwell - Null shell
from the other side. However this form can not be interpreted as a
de Sitter background in the corresponding black-hole problem. \\
\section*{Appendix: B \\ Energy Conditions }
It should be noted that the inclusion of the matter field is
acceptable  if it satisfies some physical conditions such as the
weak and dominant energy conditions. \\
{\bf1- Weak Energy Condition (WEC)} \\
The required condition for the WEC is
\begin{equation}
T_{\mu \nu}k^{\mu} k^{\nu} \geq 0
\end{equation}
where $ k^{\mu} $ be a null vector tangent to the null geodesics.
For the sake of simplicity, we consider the diagonal case of the
new solution (13), that represents null shells in the RN geometry.
The null tetrads are
\begin{eqnarray}
l^{\mu}&=& \frac{1}{\sqrt{2}}\left( \frac{ \alpha }{1- \alpha \tau
},\frac{ \alpha }{1- \alpha \tau },0,0 \right) \nonumber \\
n^{\mu}&=& \frac{1}{\sqrt{2}}\left( \frac{ \alpha }{1- \alpha \tau
},-\frac{ \alpha }{1- \alpha \tau },0,0 \right) \nonumber \\
m^{\mu}&=& - \frac{1}{\sqrt{2}} \left( 0,0, - \frac{ (1- \alpha
\tau )}{\alpha \sqrt{\Delta }}, \frac{i \alpha}{\sqrt{ \delta }
(1- \alpha \tau )} \right) \nonumber \\
\bar{m}^{\mu}&=& - \frac{1}{\sqrt{2}} \left( 0,0, - \frac{ (1-
\alpha \tau )}{\alpha \sqrt{\Delta }}, -\frac{i \alpha}{\sqrt{
\delta } (1- \alpha \tau )} \right)
\end{eqnarray}
It is found that
\begin{eqnarray}
T_{\mu \nu }l^{\mu} l^{\nu} = \Phi _{00} \nonumber \\
T_{\mu \nu }n^{\mu} n^{\nu} = \Phi _{22}
\end{eqnarray}
where
\begin{eqnarray}
\Phi_{00}&=& \frac{ \alpha ^2 (1- \alpha ^2)}{2(1- \alpha \tau
)^4} + \frac{\alpha e^{M}}{\phi} \left[ \delta (u) + M_{v} \theta
(u) \right] \nonumber \\
\Phi_{22}&=& \frac{ \alpha ^2 (1- \alpha ^2)}{2(1- \alpha \tau
)^4} + \frac{\beta e^{M}}{\phi} \left[ \delta (v) + M_{u} \theta
(v) \right]
\end{eqnarray}
which are positive and therefore, satisfies WEC. \\
{\bf2- Dominant Energy Condition (DEC)} \\
It is defined as
\begin{equation}
T^{00} \geq \left| T^{ab} \right|
\end{equation}
i.e for each $ a,b $, in the orthonormal basis the energy
dominates the other component $ T_{ab} $. The Orthonormal vectors
for  the diagonal case of the new solution are,
\begin{eqnarray}
e^{\mu}_{(0)}&=& \left( \frac{1}{\sqrt{X}},0,0,0 \right) \nonumber \\
e^{\mu}_{(1)}&=& \left( 0,-\frac{1}{\sqrt{X}},0,0 \right)\nonumber \\
e^{\mu}_{(2)}&=& \left( 0,0, \frac{1- \alpha \tau }{ \alpha
\sqrt{\Delta}},0 \right) \nonumber \\
e^{\mu}_{(3)}&=& \left( 0,0,0,-\frac{\alpha }{\sqrt{\delta } (1-
\alpha \tau )} \right)
\end{eqnarray}
The non-zero energy-momentum tensors in Orthonormal frames are
\begin{eqnarray}
T_{00}&=& \frac{1}{8 \pi}\left[ \Phi_{00} + \Phi_{22} + 2\left(
\Phi_{11}+3\Lambda \right) \right] \nonumber \\
T_{01}&=&T_{10}= \frac{1}{8 \pi} \left[ \Phi_{22} - \Phi_{00}
\right] \nonumber \\
T_{11}&=& \frac{1}{8 \pi}\left[ \Phi_{00} + \Phi_{22} - 2\left(
\Phi_{11}+3\Lambda \right) \right] \nonumber \\
T_{22}&=& \frac{1}{4 \pi}\left[ \Phi_{00} + \Phi_{11} -3\Lambda  \right] \nonumber \\
T_{33}&=& \frac{1}{4 \pi}\left[ \Phi_{11} - \Phi_{02} -3\Lambda
\right]
\end{eqnarray}
where the expressions for $ \Phi_{00}, \Phi_{22}, \Phi_{11},
\Phi_{02} $ and $ \Lambda $ is given in Appendix A.

\end{document}